\documentclass[conference]{IEEEtran}

\IEEEoverridecommandlockouts

\usepackage{cite}
\usepackage{amsmath}
\usepackage{amsopn}
\DeclareMathOperator{\diag}{diag}

\usepackage{float}

\usepackage{setspace}

\usepackage{amssymb}

\usepackage{amsfonts}

\usepackage{graphicx}

\usepackage{booktabs}

\usepackage{caption3} 
\DeclareCaptionOption{parskip}[]{} 
\usepackage[small]{caption}

\usepackage{geometry}
\usepackage{setspace,graphicx,multirow}
\usepackage{subcaption}
\usepackage{array}
\usepackage{algorithm}
\usepackage[]{algpseudocode}
\makeatletter
\def\BState{\State\hskip-\ALG@thistlm}
\makeatother
\usepackage{algcompatible}
\usepackage{url}

\usepackage{multirow}
\usepackage{hhline}
\usepackage{xspace}

\usepackage{color}

\ifodd 1
\newcommand{\com}[1]{\textbf{\color{blue} (COMMENT: #1)}} 
\else

\newcommand{\com}[1]{}
\fi

\begin{document}
\bibliographystyle{IEEEtran}
\bstctlcite{IEEEexample:BSTcontrol}
 
\title{Quasi-Static IRS: 3D Shaped Beamforming \\ for Area Coverage Enhancement}

\author{Zhenyu~Jiang,
        Xintong~Chen,
        Jiangbin~Lyu,~\IEEEmembership{Member,~IEEE},
        Liqun~Fu,~\IEEEmembership{Senior Member,~IEEE},\\
        and~Rui~Zhang,~\IEEEmembership{Fellow,~IEEE}
\thanks{Z. Jiang, X. Chen, J. Lyu and L. Fu are with the School of Informatics, Xiamen University (XMU), China, the Shenzhen Research Institute of XMU, China, and the Sichuan Institute of XMU, China. \textit{Corresponding author: Jiangbin Lyu} (email: ljb@xmu.edu.cn).

Rui Zhang is with the School of Science and Engineering, Shenzhen Research Institute of Big Data, The Chinese University of Hong Kong, China, and also with the Department of Electrical and Computer Engineering, National University of Singapore, Singapore (email: \text{rzhang@cuhk.edu.cn; elezhang@nus.edu.sg}). 
}
}

\maketitle

\begin{abstract}
Intelligent reflecting surface (IRS) is a promising paradigm to reconfigure the wireless environment for enhanced communication coverage and quality.
However, to compensate for the double pathloss effect, massive IRS elements are required, raising concerns on the scalability of cost and complexity.
This paper introduces a new architecture of \textit{quasi-static} IRS (QS-IRS), which tunes element phases via mechanical adjustment or manually re-arranging the array topology.
QS-IRS relies on massive production/assembly of purely passive elements only, and thus is suitable for ultra low-cost and large-scale deployment to enhance long-term coverage.
To achieve this end, an IRS-aided area coverage problem is formulated, which explicitly considers the element radiation pattern (ERP), with the newly introduced shape masks for the mainlobe, and the sidelobe constraints to reduce energy leakage.
An alternating optimization (AO) algorithm based on the difference-of-convex (DC) and successive convex approximation (SCA) procedure is proposed, which achieves shaped beamforming with power gains close to that of the joint optimization algorithm, but with significantly reduced computational complexity.
\end{abstract}


\IEEEpeerreviewmaketitle

\section{Introduction}




Intelligent reflecting surface (IRS), also called reconfigurable intelligent surface (RIS) or reconfigurable holographic surface (RHS), holds promise in next generation wireless networks for enhanced communication performance, which attracts extensive research on various aspects such as hardware/beamforming design, channel modelling/estimation, and network analysis/deployment (see, e.g., \cite{IRSSurver2024} and the references therein).
In particular, IRS can establish virtual line-of-sight (LoS) links\cite{ZeroOverhead} between the base station (BS) and blocked areas for improved communication quality.
Therefore, it can be deployed in indoor parking lots, hallway corners, and on building surfaces in dense urban areas for coverage enhancement.

Currently, existing literature on IRS focuses mainly on user-oriented other than region-oriented performance enhancement, where the former exploits user-specific channel state information (CSI) while the latter aims at enhancing overall area coverage with low-to-zero CSI overhead \cite{ZeroOverhead,lu2021aerial,3DBeamformingRui,lin2024broadbeam,TransparentRIS,qs_coverage,sun2024coverage}. 
An IRS-aided broadbeam design is proposed in \cite{ZeroOverhead} to illuminate the area centered around the mobile user.
A sub-array based beam broadening and flattening technique is proposed in \cite{lu2021aerial} to achieve three-dimensional (3D) beam coverage by aerial IRS.
The authors in \cite{3DBeamformingRui} improve over the approach in \cite{lu2021aerial} by employing the difference-of-convex (DC) technique\cite{rank1DCtechniqueShi} to handle the constant modulus constraint of IRS phase tuning.
A similar DC-based semi-definite programming (SDP) algorithm is proposed in \cite{lin2024broadbeam} to achieve broadbeam coverage with maximum and equal power gain within a pre-defined angular region.
The authors in \cite{TransparentRIS} propose a protocol-transparent IRS beamforming design for region-oriented coverage enhancement, and reveal that IRS should concentrate all its controllable power to the target LoS directions regardless of the Rician factors.
Narrowband systems are considered in the above works, while wideband quasi-static beam coverage is investigated in \cite{qs_coverage} by jointly optimizing transmit precoding and IRS beamforming based on statistical CSI.
User power measurement-based IRS channel estimation and reflection design for wide area coverage is recently proposed in \cite{sun2024coverage}.
Although the above works provide pioneering insights for IRS-aided broad beam design, the impact of IRS element radiation pattern (ERP)\cite{TransShiJinPathlossmodeling} is not explicitly considered, which actually plays an important role for angle-dependent 3D coverage\cite{chen2023irs}, \cite{hassan2024ERP}. 
Furthermore, due to algorithm complexity and the practical cost per IRS, evaluations are typically done with a limited number of elements (e.g., up to hundreds).
However, to compensate for the double pathloss incurred before and after reflection, massive IRS elements are required, raising concerns on the scalability of cost and complexity.

The manufacturing cost, energy consumption and reconfigurability of IRS depends on the element design and the way of assembly and phase/amplitude control.
Specifically, IRS typically consists of electronic components such as diodes, control circuits, and programmable controllers\cite{IRSHardware}, without or with active feed \cite{zhang2022active},\cite{jalali2025shape}, which enables dynamic tuning of element phase/amplitude to achieve various (real-time) beamforming functions.
However, this still requires relatively high hardware cost even for supporting a limited number of elements per printed circuit board (PCB).
In addition, channel estimation and advanced optimization algorithms are necessary to achieve dynamic beam adjustment, which leads to high system overhead and complexity.
On the other hand, traditional static reflectarrays (RAs)\cite{Yangfan2018book} typically consist of purely passive elements to achieve a given beamforming function without diodes/controllers or integrated feed, and therefore possess lower hardware cost and implementation complexity, even with zero energy consumption\cite{tmytek2025xrifle}.
However, to achieve beam coverage in different areas, RA needs to be redesigned for each specific environment and cannot be adjusted once manufactured, resulting in difficulty of mass production and poor flexibility.


Is it possible to maintain its required level of reconfigurability while significantly lowering the overall IRS manufacturing cost and energy consumption?
The answer could be positive for area coverage enhancement applications, which aim to compensate for large-scale channel effects such as pathloss and shadowing, and require IRS reconfiguration on relatively long time periods.
To this end, by leveraging the basic phase tuning mechanism of purely passive RAs \cite{Yangfan2018book},\cite{Yangfan2021hybrid} and recent advancement of electromechanical IRSs \cite{FengYJ2021metasurface},\cite{qu2023electromechanically}, we introduce a new architecture of \textit{quasi-static} IRS (QS-IRS) that tunes element phases via mechanical adjustment or manually re-arranging the array topology, thus allowing for mass production/assembly of purely passive elements.
As a result, QS-IRS is functionally similar to static RAs, and yet reconfigurable manually/mechanically, thus suitable for low-cost and large-scale deployment to enhance long-term coverage of wireless networks.

With QS-IRS, it is practically feasible to assemble massive passive elements (e.g., thousands) into panels, with the potential to achieve much higher beamforming gains compared with existing IRSs of smaller sizes.
To this end, a QS-IRS enhanced area coverage problem is formulated, which explicitly considers the impact of IRS ERP, with the newly introduced shape masks for the mainlobe, and the sidelobe constraints to reduce energy leakage.
An alternating optimization (AO) algorithm based on the DC and successive convex approximation (SCA) techniques is proposed, which achieves shaped beamforming with power gain close to that of the benchmark joint optimization algorithm, but with significantly reduced complexity. 
Furthermore, the proposed method is capable of forming non-rectangular and non-flat beam shapes that can be better tailored to the target area.
Finally, the power gain loss due to phase quantization is also evaluated.

\textit{Notations}: 
Symbols for vectors (lower case) and matrices (upper case) are in boldface. 
$\otimes$ ($\odot$) represents the Kronecker (Hadamard) product. 
$(\cdot)^T$, $(\cdot)^*$, and $(\cdot)^H$ represent the transpose, conjugate and conjugate transpose, respectively.
\text{Re}($\cdot$) takes the real part of a complex number and \text{Tr}($\cdot$) takes the matrix trace. 
$\boldsymbol{A} \succcurlyeq 0$ is a positive semi-definite (PSD) matrix.

\section{Phase-Tuning Mechanism and Design of QS-IRS}
Although the most commonly used electrically-controlled IRS can achieve millisecond-scale speed of dynamic switch, they have a large number of circuit elements (such as PIN diodes or varactors)\cite{IRSHardware} which usually incurs non-negligible insertion loss, while the complexity of the bias network significantly increases with larger IRS aperture. 
Selecting diodes with low insertion loss can alleviate this problem to some extend, but the fabrication cost will have to increase. Most importantly, electrically-controlled IRS can only realize preset beamforming functions under the condition of continuous power supply.

\begin{figure} [ht]
	\centering
	\includegraphics[width=0.9\linewidth,  trim=0 0 0 0,clip]{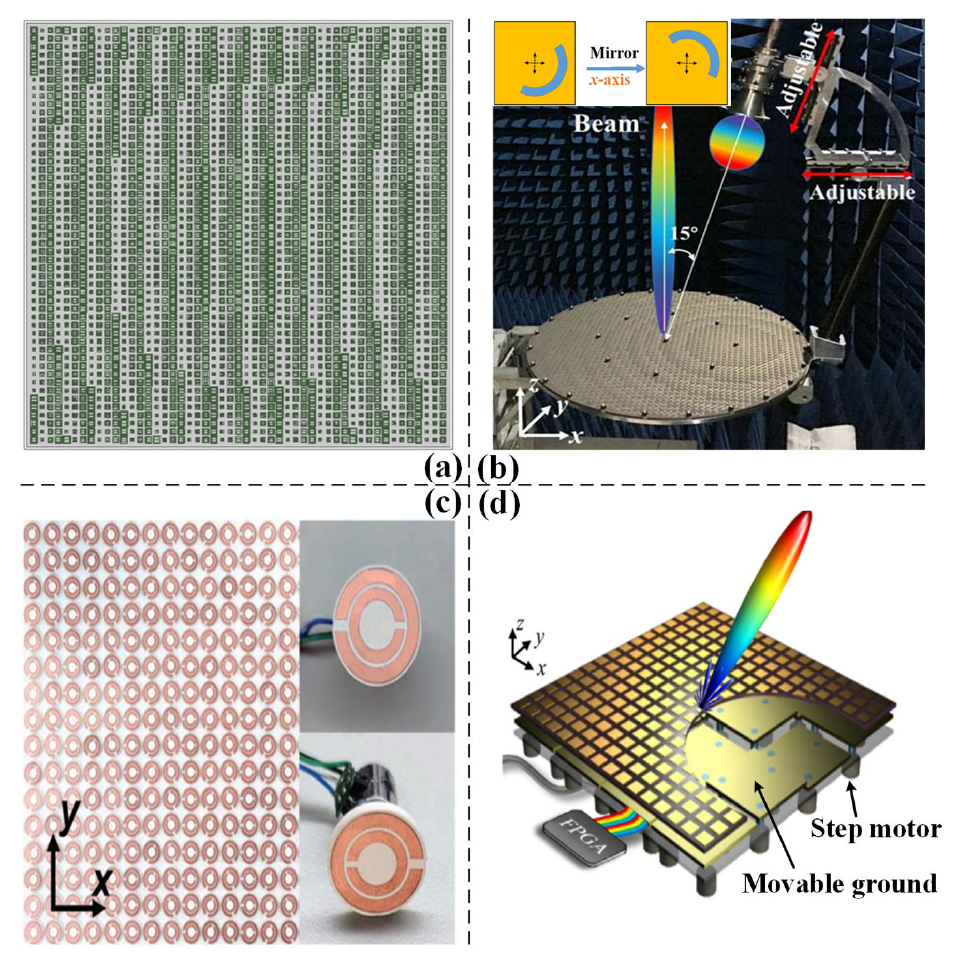}
	\caption{(a) RA manufactured by TMYTEK comprising 51$\times$51 static elements with different patch shapes/sizes\cite{tmytek2025xrifle}; (b) RA with hybrid polarization-phase tuning based on element mirroring/rotation\cite{Yangfan2021hybrid}; (c) Electromechanical IRS with rotatable patches\cite{FengYJ2021metasurface}; (d) Electromechanical IRS with movable element ground\cite{qu2023electromechanically}.\vspace{-2ex}}
    \label{Implementation}
\end{figure}

Aiming at further reducing the insertion loss, fabrication cost and power consumption, we resort to purely passive element design, for which different phase-tuning methods are available, including phase/time-delay lines, variable-size patches and variable rotation angles, etc.\cite{Yangfan2018book}.
For example, the reflection phase can be determined by the length of a delay line attached to the patch. 
In addition, elements of different patch shape/size/orientation can be designed to bear different reflection phases.
For example, \cite{tmytek2025xrifle} reported eight types of RAs to re-direct signals towards typical reflection angles.
As shown in Fig. \ref{Implementation}(a), elements of different patch shapes and sizes are systematically arranged to form an RA, without the need of external power supply.
Furthermore, RAs with hybrid-polarization phase tuning\cite{Yangfan2021hybrid} can achieve 360$^{\circ}$ phase coverage, by using the mirror combining and/or rotational combining approaches on the patch pattern design/arrangement, where an example design and its experimental verification are shown in Fig. \ref{Implementation}(b). 


The static RA enjoys low insertion loss and zero power consumption, which, however, cannot be adjusted once manufactured. 
Building on similar phase-tuning mechanisms as RAs, recently emerged electromechanical IRSs introduce reconfigurability by mechanically adjusting the element structure\cite{FengYJ2021metasurface},\cite{qu2023electromechanically}.
For example, the element orientations are controlled via step motors as in \cite{FengYJ2021metasurface}, providing high-efficiency reflection and continuous 360$^{\circ}$ reflection phases. The experiments also validate the beamforming design using a 23$\times$23-element panel, as shown in Fig. \ref{Implementation}(c), which can be extended to multi-panel and larger-scale arrays.
In addition, an electromechanical IRS with movable element ground is proposed in \cite{qu2023electromechanically} to achieve different phase shifts, as shown in Fig. \ref{Implementation}(d).
Experimental results demonstrate that it can provide stable phase configuration and high reflection efficiency (over 80$\%$).
Note that the step motors can be switched off immediately after adjustment, consuming no additional energy.
Other quasi-static designs are also available, e.g., by using manually adjustable resistors to control the element phases as in \cite{li2023resistor}, without power supply or active circuitry.

Motivated by the above phase-tuning mechanisms and designs, we introduce the new architecture of \textit{quasi-static} IRS (QS-IRS), which tunes element phases via mechanical adjustment or manually re-arranging the array topology.
QS-IRS enjoys the merits of RA including low insertion loss and zero energy consumption (after adjustment), while maintaining the required level of reconfigurability for quasi-static coverage enhancement.
The example designs\cite{FengYJ2021metasurface,qu2023electromechanically,li2023resistor} 
can be treated as different QS-IRS implementations in practice.
In this paper, we further propose a simple \textit{divide-and-assemble (DnA)} approach for QS-IRS.
Specifically, given a required phase quantization level, we first identify the unique element patterns that are equivalent via rotation or mirroring.
An illustrative example is shown in the upper right of Fig. \ref{systemmodel}, 
where patterns (a) and (c) are designed with different patch shapes to have a 90$^{\circ}$ phase difference, while their mirrored patterns (b) and (d) have 180$^{\circ}$ difference with them, respectively, thus providing 2-bit quantization\cite{Yangfan2021hybrid}.
As a result, we identify two distinct patterns (a) and (c) for mass production.
Second, according to the desired phase distribution for each target area, we can manually assemble the (properly rotated) elements into panels\footnote{The elements can be pasted directly on smooth surface or on acrylic plates.} to form an ultra-large QS-IRS. 
Based on the above DnA approach, we identify as few patterns as possible for mass production and assembly, thereby significantly reducing the manufacturing cost/failure rate per element (instead of per IRS), along with cost/energy savings for diodes/controllers/bias networks/step motors, etc.
Note that the proposed QS-IRS can be regarded as a manually reconfigurable RA, which inherits both the phase-tuning mechanisms\cite{Yangfan2018book},\cite{Yangfan2021hybrid} and the design principles of mechanically assembling elements into panels \cite{FengYJ2021metasurface},\cite{qu2023electromechanically},\footnote{Experiments in Fig. \ref{Implementation}(c) and Fig. \ref{Implementation}(d) are tested at sub-6 GHz bands.
Our proposed DnA approach can be extended to higher frequency bands with smaller element sizes, by designing modular subarray patterns of appropriate sizes for assembly, which is left for our future work.} for further reducing the gross cost per IRS.

\section{System Model}

\begin{figure} [t]
	\centering
	\includegraphics[width=0.9\linewidth,  trim=0 0 0 10,clip]{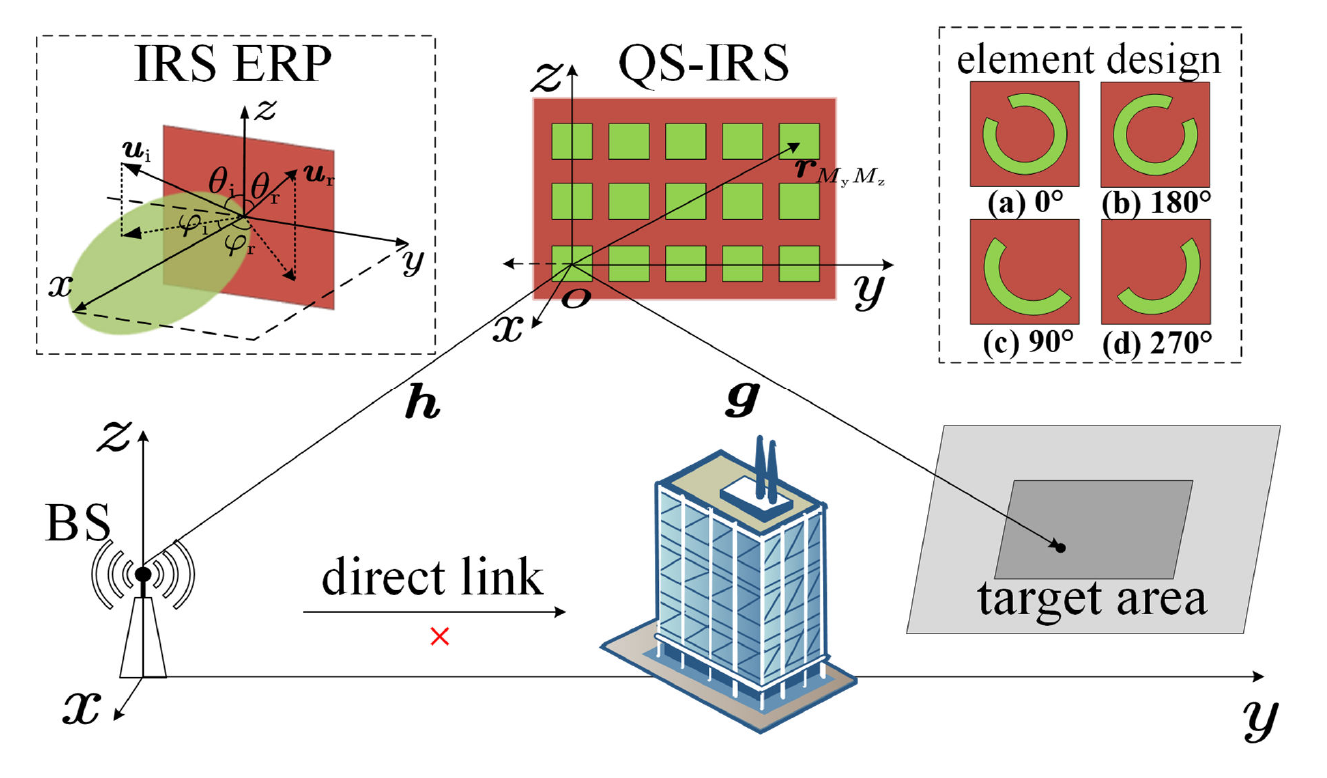}
	\caption{Area coverage aided by QS-IRS beamforming.\vspace{-3ex}}\label{systemmodel}
\end{figure}

Consider far-field communications between the BS and a target area on the ground with severely blocked direct links.
We focus on practical scenarios where a QS-IRS is appropriately deployed\footnote{The case with multiple distributed IRSs\cite{xu2024distributed} is left for future investigation.} such that there exists LoS paths from the BS to the target area via the QS-IRS, as in \cite{ZeroOverhead},\cite{3DBeamformingRui}.\footnote{The non-LoS scenario with non-negligible direct links will be considered in our future work, whereby statistical analysis of link power and outage probability under (e.g., Rician) fading channel can be incorporated in our system model and optimization framework.}
Consider one omnidirectional antenna at the receiver, and one (synthesized) directional antenna (beam) at the BS with a peak power gain of $G_\text{t}$ pointing towards the QS-IRS.\footnote{Our proposed method can be fully protocol-transparent as in \cite{TransparentRIS}, under given BS antenna configuration.}
Assume that $M\triangleq M_{\text{y}}\times M_{\text{z}}$ IRS elements are arranged as a uniform planar array (UPA) with $M_{\text{y}}$ and $M_{\text{z}}$ elements along the $y\text{-}$ and $z\text{-}$axis, respectively, and with the first element at the origin, as shown in Fig. \ref{systemmodel}.
Consider the IRS ERP model \cite{TransShiJinPathlossmodeling} given by $G F(\Pi)$, where $G$ denotes the peak power gain and $F(\Pi)$ denotes the normalized power pattern from/to the azimuth and elevation angles $\Pi\triangleq (\varphi,\theta)$, i.e.,
\begin{equation}\small
    F(\Pi) \triangleq
     \left\{
        \begin{array}{cc}
        (\sin \theta \cos \varphi)^{\frac{G}{2}-1}, & \theta \in [0,\pi], \varphi \in [-\frac{\pi}{2},\frac{\pi}{2}], \\
        0, &  {\rm otherwise}.
        \end{array}
    \right.
\end{equation}

\subsection{Array Steering Vector}
Denote $\boldsymbol{u}_\text{i} \triangleq [\cos\varphi_{\text{i}} \sin\theta_{\text{i}},\sin\varphi_{\text{i}} \sin\theta_{\text{i}},\cos\theta_{\text{i}}]^T$ as the unit direction vector of the BS position w.r.t. the origin, with $\Pi_\text{i} \triangleq (\varphi_\text{i}, \theta_\text{i})$ denoting the corresponding (horizontal, vertical) angle.
The components of the steering vector incident on the QS-IRS along the $y\text{-}$ and $z\text{-}$axis can be expressed as
\begin{align} \label{receiveSteeringV}
\boldsymbol{a}_{\text{i}\text{y}}(\Pi_\text{i}) \triangleq [1, \cdots, e^{j\frac{2\pi f_c}{c} (\boldsymbol{r}_{M_{\text{y}}} \cdot \boldsymbol{u}_\text{i})}]^T \in \mathbb{C}^{M_{\text{y}} \times 1}, \\
\boldsymbol{a}_{\text{i}\text{z}}(\Pi_\text{i}) \triangleq [1, \cdots, e^{j\frac{2\pi f_c}{c} (\boldsymbol{r}_{M_{\text{z}}} \cdot \boldsymbol{u}_\text{i})}]^T \in \mathbb{C}^{M_{\text{z}} \times 1},
\end{align}%
where $\boldsymbol{r}_{m_{\text{y}}} \triangleq [0,(m_{\text{y}}-1)d_{\text{y}},0]^T$ and $\boldsymbol{r}_{m_{\text{z}}} \triangleq [0,0,(m_{\text{z}}-1)d_{\text{y}}]^T$ denote the position vectors of the $(m_{\text{y}},m_{\text{z}})$-th element along the $y\text{-}$ and $z\text{-}$axis, with $d_{\text{y}}$ and $d_{\text{z}}$ being the element spacing;
$c$ denotes the speed of light and $f_\text{c}$ denotes the carrier frequency.
Then, the incident steering vector is given by $\boldsymbol{a}_\text{i}(\Pi_\text{i}) \triangleq \boldsymbol{a}_{\text{i}\text{y}}(\Pi_\text{i}) \otimes \boldsymbol{a}_{\text{i}\text{z}}(\Pi_\text{i})$.
Similarly, the components of the reflect steering vector are given by 
\begin{align} \label{reflectSteeringV}
\boldsymbol{a}_{\text{r}\text{y}}(\Pi_\text{r}) \triangleq [1, \cdots, e^{j\frac{2\pi f_c}{c} (\boldsymbol{r}_{M_{\text{y}}} \cdot \boldsymbol{u}_\text{r})}]^T \in \mathbb{C}^{M_{\text{y}} \times 1}, \\
\boldsymbol{a}_{\text{r}\text{z}}(\Pi_\text{r}) \triangleq [1, \cdots, e^{j\frac{2\pi f_c}{c} (\boldsymbol{r}_{M_{\text{z}}} \cdot \boldsymbol{u}_\text{r})}]^T \in \mathbb{C}^{M_{\text{z}} \times 1},
\end{align}%
where $\Pi_\text{r} \triangleq (\varphi_\text{r}, \theta_\text{r})$ denotes the reflect angle, and $\boldsymbol{u}_\text{r} \triangleq [\cos\varphi_{\text{r}} \sin\theta_{\text{r}},\sin\varphi_{\text{r}} \sin\theta_{\text{r}},\cos\theta_{\text{r}}]^T$ denotes the unit direction vector of the UE position.
The reflect steering vector of the QS-IRS is given by $\boldsymbol{a}_\text{r}(\Pi_\text{r}) \triangleq \boldsymbol{a}_{\text{r}\text{y}}(\Pi_\text{r}) \otimes \boldsymbol{a}_{\text{r}\text{z}}(\Pi_\text{r})$.
Therefore, the full steering vector of the QS-IRS is given as 
\begin{align} \label{FullSteeringV}
    & \boldsymbol{a}(\Pi_\text{i},\Pi_\text{r}) = \boldsymbol{a}_\text{i}(\Pi_\text{i}) \odot \boldsymbol{a}_\text{r}(\Pi_\text{r}) \notag \\
    & = \left( \boldsymbol{a}_{\text{i}\text{y}}(\Pi_\text{i}) \otimes \boldsymbol{a}_{\text{i}\text{z}}(\Pi_\text{i}) \right) \odot \left( \boldsymbol{a}_{\text{r}\text{y}}(\Pi_\text{r}) \otimes \boldsymbol{a}_{\text{r}\text{z}}(\Pi_\text{r}) \right) \notag \\
    & = \left( \boldsymbol{a}_{\text{i}\text{y}}(\Pi_\text{i}) \odot \boldsymbol{a}_{\text{r}\text{y}}(\Pi_\text{r}) \right) \otimes \left( \boldsymbol{a}_{\text{i}\text{z}}(\Pi_\text{i}) \odot \boldsymbol{a}_{\text{r}\text{z}}(\Pi_\text{r}) \right) \notag \\
    & = \boldsymbol{a}_\text{y}(\Pi_\text{i},\Pi_\text{r}) \otimes \boldsymbol{a}_\text{z}(\Pi_\text{i},\Pi_\text{r}),
\end{align}
where $\boldsymbol{a}_\text{y}(\Pi_\text{i},\Pi_\text{r})$ and $\boldsymbol{a}_\text{z}(\Pi_\text{i},\Pi_\text{r})$ represent the $y$- and $z$- components of the full steering vector, respectively.

\subsection{Channel Power Gain}
The BS-to-IRS channel can then be expressed as
\begin{equation}
    \boldsymbol{h} \triangleq \eta_1(\Pi_\text{i}) \beta_1 \boldsymbol{a}_\text{i}(\Pi_\text{i})\in \mathbb{C}^{M \times 1},
\end{equation}
where $\eta_1(\Pi_\text{i}) \triangleq \sqrt{G_\text{t} G F(\Pi_\text{i})}$ and $\beta_1 \triangleq \frac{c}{4\pi d_1 f_c} e^{-j \frac{2\pi d_1 f_c}{c}}$, with $d_1$ being the BS-IRS distance. 
Similarly, the IRS-to-UE channel is given by
\begin{equation}
\boldsymbol{g}\triangleq\eta_2(\Pi_\text{r}) \beta_2 \boldsymbol{a}_\text{r}(\Pi_\text{r})\in \mathbb{C}^{M \times 1},
\end{equation}
where $\eta_2(\Pi_\text{r}) \triangleq \sqrt{G F(\Pi_\text{r})}$ and $\beta_2 \triangleq \frac{c}{4\pi d_2 f_c} e^{-j \frac{2\pi d_2 f_c}{c}}$, with $d_2$ being the IRS-UE distance.
Denote the phase compensation (or beamforming) vector of the QS-IRS as $\boldsymbol{w} \triangleq \left[ e^{j\phi_{11}},e^{j\phi_{12}},\cdots,e^{j\phi_{M_{\text{y}}M_{\text{z}}}} \right]^T \in \mathbb{C}^{M \times 1}$, and its diagonal matrix version as $\boldsymbol{\Omega} \triangleq \diag (\boldsymbol{w})$. 
Then the cascaded BS-IRS-receiver channel is given by
\begin{align} \label{cascadedchannel}
    & \boldsymbol{g}^T \boldsymbol{\Omega} \boldsymbol{h} =\eta(\Pi_\text{i},\Pi_\text{r}) \beta \boldsymbol{a}^T(\Pi_\text{i},\Pi_\text{r}) \boldsymbol{w}, 
\end{align} 
where $\eta(\Pi_\text{i},\Pi_\text{r}) \triangleq \eta_1(\Pi_\text{i}) \eta_2(\Pi_\text{r})$ and $\beta  \triangleq \beta_1 \beta_2$.
As a result, the cascaded channel power gain (including antenna and IRS ERP gains) normalized to the overall pathloss $\beta^2$ is given by
\begin{align}
&\gamma(\Pi_\text{i},\Pi_\text{r}) 
\triangleq \left|\boldsymbol{g}^T \boldsymbol{\Omega} \boldsymbol{h} \right|^2/ \beta^2 \notag \\ 
& = \eta^2(\Pi_\text{i},\Pi_\text{r}) \left| \boldsymbol{a}^T(\Pi_\text{i},\Pi_\text{r}) \boldsymbol{w} \right|^2 \label{eq:channelPowerGain_a} \\ 
& = \eta^2(\Pi_\text{i},\Pi_\text{r}) \left| \boldsymbol{a}^T_\text{y}(\Pi_\text{i},\Pi_\text{r}) \boldsymbol{w}_\text{y} \right|^2 \left| \boldsymbol{a}^T_\text{z}(\Pi_\text{i},\Pi_\text{r}) \boldsymbol{w}_\text{z} \right|^2, \label{eq:channelPowerGain_b}
\end{align}
where \eqref{eq:channelPowerGain_b} is due to \eqref{FullSteeringV} and the decomposition of $\boldsymbol{w}=\boldsymbol{w}_\text{y} \otimes\boldsymbol{w}_\text{z}$, with $y\text{-}$ component $\boldsymbol{w}_\text{y}\triangleq \left[ e^{j\phi_{1}},\cdots,e^{j\phi_{M_{\text{y}}}} \right]^T$ and $z\text{-}$ component $\boldsymbol{w}_\text{z}\triangleq \left[ e^{j\phi_{1}},\cdots,e^{j\phi_{M_{\text{z}}}} \right]^T$.

\section{Problem Formulation and Solution}
Due to the potentially different gain requirements at various positions within the target area, this paper aims to maximize the shaped beamforming gain towards the target area, by optimizing the reflection phases of the QS-IRS.

\subsection{Joint Optimization}
To provide shaped beam coverage, the IRS reflect angular range is divided into the mainlobe and sidelobe zones, each represented by a set of discrete angles, i.e., $\Pi_p = (\varphi_p, \theta_p), p\in \mathbf{P}$ for the mainlobe, and $\Pi_q = (\varphi_q, \theta_q), q\in \mathbf{Q}$ for the sidelobes.
Moreover, we introduce a given shape mask $d(\Pi_p)$ for the mainlobe to impose finer shape requirements.
As a result, we aim to maximize the common power gain $\rho$ in the mainlobe subject to a given gain gap $\delta$ over the sidelobes, i.e.,
\begin{align} 
\text{(P1)} &\max_{\boldsymbol{w}, \rho} \,
 10\log_{10}\rho \notag\\ 
\mathrm{s.t.} \,
& \gamma(\Pi_\text{i},\Pi_p) \ge \rho d(\Pi_p), p\in \mathbf{P}, \quad \label{mainlodeconstr} \\ 
& \gamma(\Pi_\text{i},\Pi_q) \leq \rho / \delta, q\in \mathbf{Q}, \quad \label{sidelodeconstr} \\ 
& \left| \boldsymbol{w}[m] \right| = 1, m = 1,\cdots,M, \label{modulus1} \quad 
\end{align} 
where \eqref{mainlodeconstr} and \eqref{sidelodeconstr} are the mainlobe and sidelobe constraints, respectively, and \eqref{modulus1} represents the constant modulus constraint of IRS phase control. 
Note that the above problem formulation differs from that in \cite{3DBeamformingRui} by the objective of shaped beam coverage, the consideration of angle-dependent antenna patterns $\eta^2(\Pi_\text{i},\Pi_\text{r})$, and the newly introduced shape constraints \eqref{mainlodeconstr} and \eqref{sidelodeconstr}.
In particular, by explicitly enforcing the mainlobe-to-sidelobe gap, the mainlobe gain could be further improved compared to the case without \eqref{sidelodeconstr} (as in \cite{3DBeamformingRui}), as illustrated later in Section \ref{Results}.
However, the angle-dependent constraints \eqref{mainlodeconstr} and \eqref{sidelodeconstr} entangle with each other (rendering the proposed decomposition in \cite{3DBeamformingRui} inapplicable) as well as the non-convex constraint \eqref{modulus1}, making (P1) challenging to solve.
Here we first apply the DC technique \cite{rank1DCtechniqueShi} to tackle the constraint \eqref{modulus1}, by optimizing the overall IRS phase vector $\boldsymbol{w}$ in \eqref{eq:channelPowerGain_a}. 

Specifically, to tackle the non-convex constraints \eqref{mainlodeconstr} and \eqref{modulus1}, the channel power gain in \eqref{eq:channelPowerGain_a} is written as
\begin{align} \label{centralizedTrace}
    & \gamma(\Pi_\text{i},\Pi_\text{r})=\eta^2(\Pi_\text{i},\Pi_\text{r}) \left| \boldsymbol{a}^T(\Pi_\text{i},\Pi_\text{r}) \boldsymbol{w} \right|^2 \notag \\
    &= \eta^2(\Pi_\text{i},\Pi_\text{r}) \boldsymbol{w}^H \boldsymbol{a}^{*}(\Pi_\text{i},\Pi_\text{r}) \boldsymbol{a}^T(\Pi_\text{i},\Pi_\text{r}) \boldsymbol{w} \notag \\
    &= \eta^2(\Pi_\text{i},\Pi_\text{r}) \text{Re}\left( \text{Tr}(\boldsymbol{A}(\Pi_\text{i},\Pi_\text{r}) \boldsymbol{W}) \right),
\end{align}
which becomes linear in the new matrix variable $\boldsymbol{W} \triangleq \boldsymbol{w} \boldsymbol{w}^H\in \mathbb{C}^{M \times M}$,
with coefficient matrix $\boldsymbol{A}(\Pi_\text{i},\Pi_\text{r}) \triangleq \boldsymbol{a}^{*}(\Pi_\text{i},\Pi_\text{r}) \boldsymbol{a}^T(\Pi_\text{i},\Pi_\text{r})$.
As a result, both \eqref{mainlodeconstr} and \eqref{sidelodeconstr} are converted into linear constraints, i.e.,
\begin{align}
   & \eta^2(\Pi_\text{i},\Pi_p) \text{Re}\left( \text{Tr}(\boldsymbol{A}(\Pi_\text{i},\Pi_p) \boldsymbol{W}) \right) \ge \rho d(\Pi_p), \label{mainlodeinequ_centralized} \\ 
    & \eta^2(\Pi_\text{i},\Pi_q) \text{Re}\left( \text{Tr}(\boldsymbol{A}(\Pi_\text{i},\Pi_q) \boldsymbol{W}) \right) \leq \rho / \delta,\label{sidelodeinequ_centralized} 
\end{align}
while the constant modulus constraint \eqref{modulus1} is equivalent to
\begin{align}
    \label{diagone} \boldsymbol{W}[m,m] =& 1, m = 1,\cdots,M, \\
    \label{rank1} \text{rank}(\boldsymbol{W}) =& 1.
\end{align}

Note that the rank constraint \eqref{rank1} is non-convex and non-continuous, which is further tackled by the DC technique\cite{rank1DCtechniqueShi}.
Specifically, for a PSD matrix $\boldsymbol{W}$ with $\text{Tr}(\boldsymbol{W}) >0$, we have
\begin{equation}\label{DCconstr}
    \text{rank}(\boldsymbol{W}) = 1 \Longleftrightarrow \lVert \boldsymbol{W} \rVert_{*} - \lVert \boldsymbol{W} \rVert_{2} = 0,
\end{equation}
where $\lVert \boldsymbol{W} \rVert_{*}$ and $\lVert \boldsymbol{W} \rVert_{2}$ represent the nuclear norm and the spectral norm, respectively, which are convex functions of $\boldsymbol{W}$.
However, the difference of these two convex functions is non-convex.
To satisfy the DC constraint \eqref{DCconstr}, the DC term is added as a penalty in the objective with weight $\sigma$, i.e.,
\begin{equation}\label{objectivepenalty}
    \max_{\boldsymbol{W}, \rho} \,\, 10\log_{10}\rho - \sigma (\lVert \boldsymbol{W} \rVert_{*} - \lVert \boldsymbol{W} \rVert_{2}),
\end{equation}
which is then iteratively solved using the SCA method\cite{3DBeamformingRui}.
Specifically, in the $n\text{-}$th iteration, $\lVert \boldsymbol{W} \rVert_{2}$ is approximated by its first-order Taylor expansion at a given local point $\boldsymbol{W}^n$, i.e.,
$\lVert \boldsymbol{W}^n \rVert_{2}  + \text{Re} \big( \text{Tr} \big( \left( \partial_{\boldsymbol{W}^n} \lVert \boldsymbol{W} \rVert_2 \right) ( \boldsymbol{W} - \boldsymbol{W}^n ) \big)\big)$, 
where the sub-gradient of $\lVert \boldsymbol{W} \rVert_{2}$ at $\boldsymbol{W}^n$ can be calculated as $\partial_{\boldsymbol{W}^n} \lVert \boldsymbol{W} \rVert_2 = \boldsymbol{v} \boldsymbol{v}^H$, with $\boldsymbol{v}$ being the dominant singular vector of $\boldsymbol{W}^n$.
Then, each iteration solves a convex subproblem, i.e.,
\begin{align*}
\text{(P2)} \max_{\boldsymbol{W}, \rho} \quad 
& 10\log_{10}\rho - \sigma \left( \lVert \boldsymbol{W} \rVert_{*} - \left( \lVert \boldsymbol{W}^n \rVert_{2} \right. \right. \\
     & \left. \left. + \text{Re} \big( \text{Tr} \big( \left( \partial_{\boldsymbol{W}^n} \lVert \boldsymbol{W} \rVert_2 \right) ( \boldsymbol{W} - \boldsymbol{W}^n ) \big) \big) \right) \right),\\
\mathrm{s.t.} \quad & \eqref{mainlodeinequ_centralized}, \eqref{sidelodeinequ_centralized}, \eqref{diagone}, \textrm{and }\boldsymbol{W} \succcurlyeq 0.
\end{align*}
The optimal solution $\boldsymbol{W}^*$ to (P2) can be found by the CVX toolbox, which is then taken as the local point $\boldsymbol{W}^{n+1}$ in the next iteration. 
The iteration continues until the increase of the objective value is below a certain threshold $\xi$.
Eventually, the beamforming vector $\boldsymbol{w}$ is obtained as the dominant singular vector of $\boldsymbol{W}$ with the largest singular value.

Note that solving (P2) using the interior-point method is of complexity $\mathcal{O}(M^{6.5})$\cite{3DBeamformingRui}, which is suitable for small/moderate $M$.
However, for large-scale QS-IRS (e.g., with $M>1000$),
the computational complexity becomes prohibitive.
An initial attempt to reduce complexity is to decompose the power gain in \eqref{eq:channelPowerGain_b} into the sum of $y$- and $z$- components in dB scale, which transforms \eqref{mainlodeconstr} into convex constraints, but turning \eqref{sidelodeconstr} into non-convex. 

\subsection{Alternating Optimization (AO)}
To this end, the power gain in \eqref{eq:channelPowerGain_b} is first written as
\begin{small} 
$\eta^2(\Pi_\text{i},\Pi_\text{r}) \text{Re}(\text{Tr}(\boldsymbol{A}_\text{y}(\Pi_\text{i},\Pi_\text{r}) \boldsymbol{W}_\text{y})) \text{Re}(\text{Tr}(\boldsymbol{A}_\text{z}(\Pi_\text{i},\Pi_\text{r}) \boldsymbol{W}_\text{z}))$
\end{small},
similarly to \eqref{centralizedTrace}, 
where $\boldsymbol{W}_\text{y} \triangleq \boldsymbol{w}_\text{y} \boldsymbol{w}_\text{y}^H$ and $\boldsymbol{W}_\text{z} \triangleq \boldsymbol{w}_\text{z} \boldsymbol{w}_\text{z}^H$ are both rank-one PSD matrices.
In this way, the power gain $\gamma(\Pi_\text{i},\Pi_\text{r})$ becomes the product of two functions affine in $\boldsymbol{W}_\text{y}$ and $\boldsymbol{W}_\text{z}$, respectively, which is not jointly convex but is convex in each one alone.
Therefore, we apply the AO strategy by optimizing $\boldsymbol{W}_\text{y}$ with given $\boldsymbol{W}_\text{z}$ (and vice versa), to maximize $10\log_{10}\rho - \sigma\big( (\lVert \boldsymbol{W}_\text{y} \rVert_{*} - \lVert \boldsymbol{W}_\text{y} \rVert_{2})+(\lVert \boldsymbol{W}_\text{z} \rVert_{*} - \lVert \boldsymbol{W}_\text{z} \rVert_{2})\big)$.
Following the DC-SCA procedure, we first solve
\begin{align*}
\text{(P3)} \max_{\boldsymbol{W}_\text{y}, \rho} \quad & 10\log_{10}\rho - \sigma \left( \lVert \boldsymbol{W}_\text{y} \rVert_{*} - \left( \lVert \boldsymbol{W}_\text{y}^n \rVert_{2} \right. \right. \\
     & \left. \left. + \text{Re} \big( \text{Tr} \big( ( \partial_{\boldsymbol{W}_\text{y}^n} \lVert \boldsymbol{W}_\text{y} \rVert_2 ) ( \boldsymbol{W}_\text{y} - \boldsymbol{W}_\text{y}^n ) \big) \big) \right) \right),\\
\mathrm{s.t.} \quad & \eqref{mainlodeconstr}; \eqref{sidelodeconstr}; \boldsymbol{W}_\text{y}[m_\text{y},m_\text{y}] = 1,\forall m_\text{y}; \textrm{and }\boldsymbol{W}_\text{y} \succcurlyeq 0.
\end{align*}
The obtained solution $\boldsymbol{W}_\text{y}$ is then taken as the fixed value used in the next subproblem of optimizing $\boldsymbol{W}_\text{z}$.
Such AO procedure continues until the threshold $\xi$ or a maximum number of iterations $\zeta$ is met, with details summarized in Algorithm 1.
Moreover, $\boldsymbol{W}_\text{z}$ and $\boldsymbol{W}_\text{y}$ are initialized to align the IRS element phases to focus on the area center.

Solving each convex subproblem (P3) using the interior-point method is of complexity $\mathcal{O}(M_{\text{y}}^{6.5})$ or $\mathcal{O}(M_{\text{z}}^{6.5})$ based on similar analysis as in \cite{3DBeamformingRui}.
Meanwhile, the DC-SCA procedure in $y/z$-axis will converge to a stationary point, with the DC component strictly decreasing and convergent according to Proposition 3 in \cite{rank1DCtechniqueShi}. 
As a result, the objective value will be non-decreasing as the number of AO iterations increases until convergence, as will be illustrated later in Fig. \ref{ConvergenceAnalysis}.



\begin{algorithm} [t]
\textsl{}\setstretch{0.7}
\renewcommand{\algorithmicrequire}{\textbf{Input:}}
\renewcommand{\algorithmicensure}{\textbf{Output:}}
	\caption{AO algorithm for 3D shaped beamforming}
	\begin{algorithmic}[1]
		\REQUIRE $d(\Pi_p)$, $\mathbf{P}$, $\mathbf{Q}$, $\delta$, $\sigma$, $\xi$.
            \STATE Initialization: $\zeta \gets 0$, $\boldsymbol{W}_\text{z}$.
            \WHILE {the maximum number of iterations is less than $\zeta$}
            \STATE Initialization: $n \gets 0$, $\boldsymbol{W}_\text{y}^{0}$.
		\WHILE {the increase of objective value is above $\xi$}
		\STATE Obtain $\partial_{\boldsymbol{W}_\text{y}^n} \lVert \boldsymbol{W}_\text{y} \rVert_2$ via SVD.
            \STATE Solve (P3) using CVX and obtain $\boldsymbol{W}_\text{y}^{n+1}$.
            \STATE $n \gets n+1$.
            \ENDWHILE
            \STATE The solution $\boldsymbol{W}_\text{y}$ is taken as given value.
            \STATE Obtain $\boldsymbol{W}_\text{z}$ via similar procedures like steps 4$\sim$8.
            \STATE The solution $\boldsymbol{W}_\text{z}$ is taken as given value.
            \STATE $\zeta \gets \zeta+1$.
            \ENDWHILE
             \STATE Obtain $\boldsymbol{w}_\text{y}^{opt}$ and $\boldsymbol{w}_\text{z}^{opt}$ via SVD on $\boldsymbol{W}_\text{y}$ and $\boldsymbol{W}_\text{z}$.
		\ENSURE  
  QS-IRS beamforming vector $\boldsymbol{w}^{opt} = \boldsymbol{w}_\text{y}^{opt} \otimes \boldsymbol{w}_\text{z}^{opt}$.
	\end{algorithmic}  
\end{algorithm}

\section{Numerical Results} \label{Results}
Numerical results are provided to evaluate the performance of our proposed shaped beamforming design. 
The following parameters are used if not mentioned otherwise: 
$f_\text{c}=3.5$ GHz, $G_\text{t} = 14.5$ dB, $G = 4$ dB, $\delta = 10$ dB, $\sigma = 20$, $\xi= 0.001$, $\zeta = 10$, $\varphi_\text{i} = -45^{\circ}$, $\theta_\text{i} = 144^{\circ}$, $\varphi_\text{r} \in [-90^{\circ},90^{\circ}]$, $\theta_\text{r} \in [90^{\circ},180^{\circ}]$, $d_\text{y} = d_\text{z} = \frac{c}{2f_\text{c}}$, and $M_\text{y} = M_\text{z} = 48$.
$\mathbf{P}$ contains the index of angles with $\varphi_p \in [-15^{\circ},15^{\circ}]$ and $\theta_p \in [110^{\circ},140^{\circ}]$, sampled per $10^{\circ}$.
To allow for smooth round-offs between mainlobe and sidelobes, $\mathbf{P}$ is surrounded by a $10^{\circ}$ gap region, with the rest of angles included in $\mathbf{Q}$.
The simulations were run in MATLAB using single-core 2.1 GHz CPU and 64 GB memory.

\subsection{3D Shaped Beamforming} 
\begin{figure}
	\centering
	\includegraphics[width=1.0\linewidth,  trim=0 0 0 0,clip]{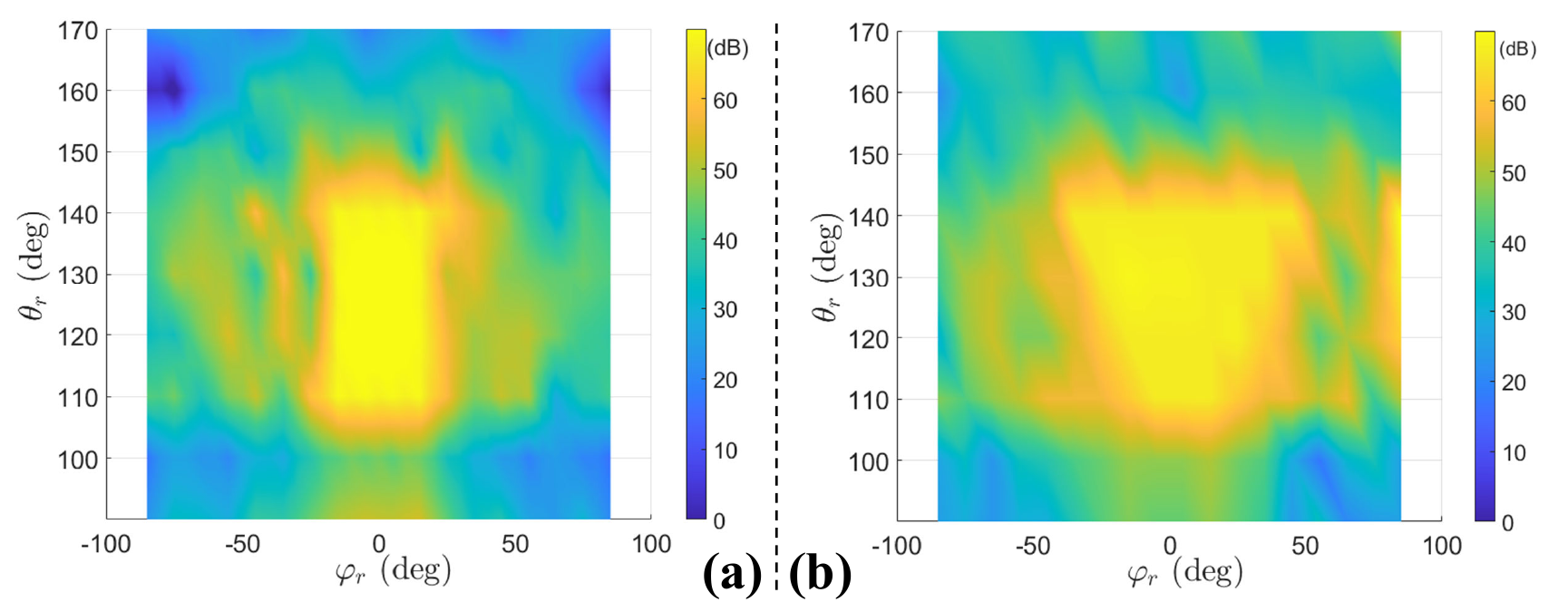}
	\caption{(a) Square shape coverage; (b) Trapezoidal shape coverage.\vspace{-2ex}}
    \label{Coverage}
\end{figure}

First, the shaped beam realized by our AO algorithm is illustrated in Fig. \ref{Coverage}(a) using flat-top beams, i.e., $d(\Pi_p) = 1$, $p\in \mathbf{P}$, with a square or trapezoidal shape, respectively.
It can be seen that the proposed algorithm is capable of generating beams with different shaped regions, with more than 10 dB power gain over sidelobes.
Note that the method in \cite{3DBeamformingRui} decomposes the 2D flat-top beamforming into two independent 1D array beamforming, which, due to symmetry, cannot achieve beam shapes such as the trapezoidal shape.
In contrast, our AO procedure allows explicit consideration of 3D ERP and also the correlation along the two array directions, thus enabling tailored beam shapes, as shown in Fig. \ref{Coverage}(b).

\begin{figure} [htb]
	\centering
\includegraphics[width=0.65\linewidth,  trim=0 0 0 0,clip]{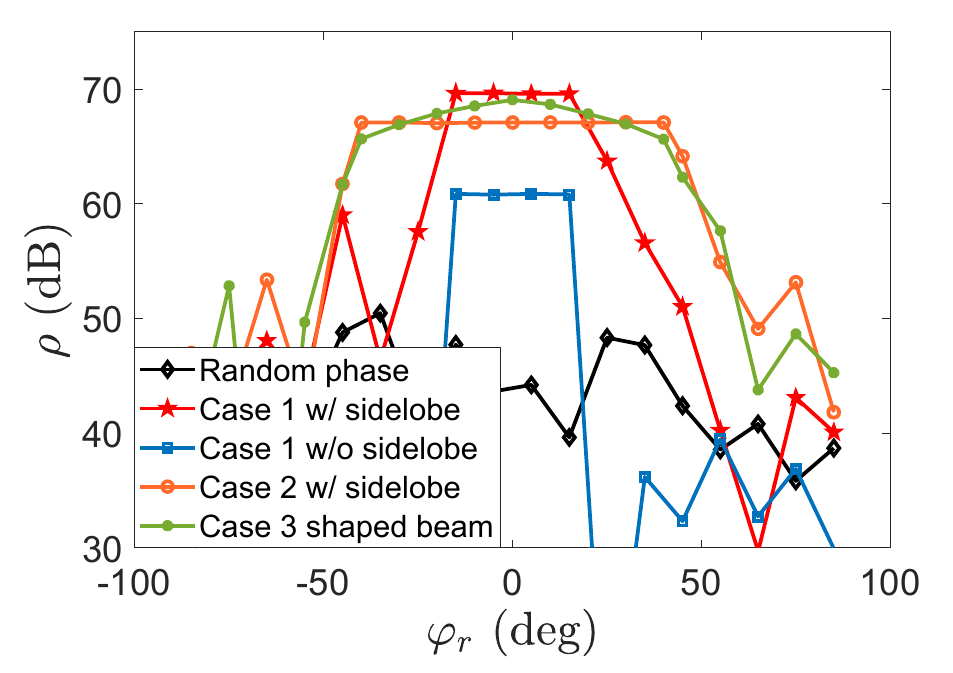}
	\caption{Power gain performance under difference schemes and scenarios with $M_\text{y} = M_\text{z} = 48$, observed at $\theta_\text{r} = 140^{\circ}$.\vspace{-2ex}}\label{Necessity}
\end{figure}

More results are plotted in Fig. \ref{Necessity} to compare the power gain performance under different schemes and scenarios.
For case 1 with the default beamwidth, our design explicitly considers the sidelobe constraints \eqref{sidelodeconstr}, which achieves gain improvement over the case in \cite{3DBeamformingRui} without \eqref{sidelodeconstr}.
This could be due to the help of sidelobe constraints to reduce energy leakage in sidelobes and thus raise energy concentration in the mainlobe.
For case 2 with a wider beamwidth $\varphi_p \in [-40^{\circ},40^{\circ}]$ and $\theta_p \in [120^{\circ},140^{\circ}]$, the gain is lower than that in case 1, similarly due to energy conservation.
For case 3, we demonstrate another (parabolically) shaped beam pattern as in 3GPP TR 38.901, given in dB scale as
\begin{align}
   F(\Pi_\text{r}) = -L \left( \frac{\Pi_\text{r}-\Pi_0}{\Delta\Pi_{3\text{dB}}} \right)^2, \lVert \Pi_\text{r} - \Pi_0 \lVert \leq \Delta\Pi_{3\text{dB}},
\end{align}
where $L=3$; $\Delta\Pi_{3\text{dB}}$ represents one-half of the half-power beamwidth (HPBW); and $\Pi_0$ represents the boresight of the beam.
It can be seen that the gain achieves its peak in the boresight direction and gradually decreases within the HPBW.
Finally, all the above gains are significantly higher than that of the baseline case with random IRS phases.

\vspace{-3ex}

\begin{figure} [htb]
	\centering
	\includegraphics[width=0.95\linewidth,  trim=0 0 0 5,clip]{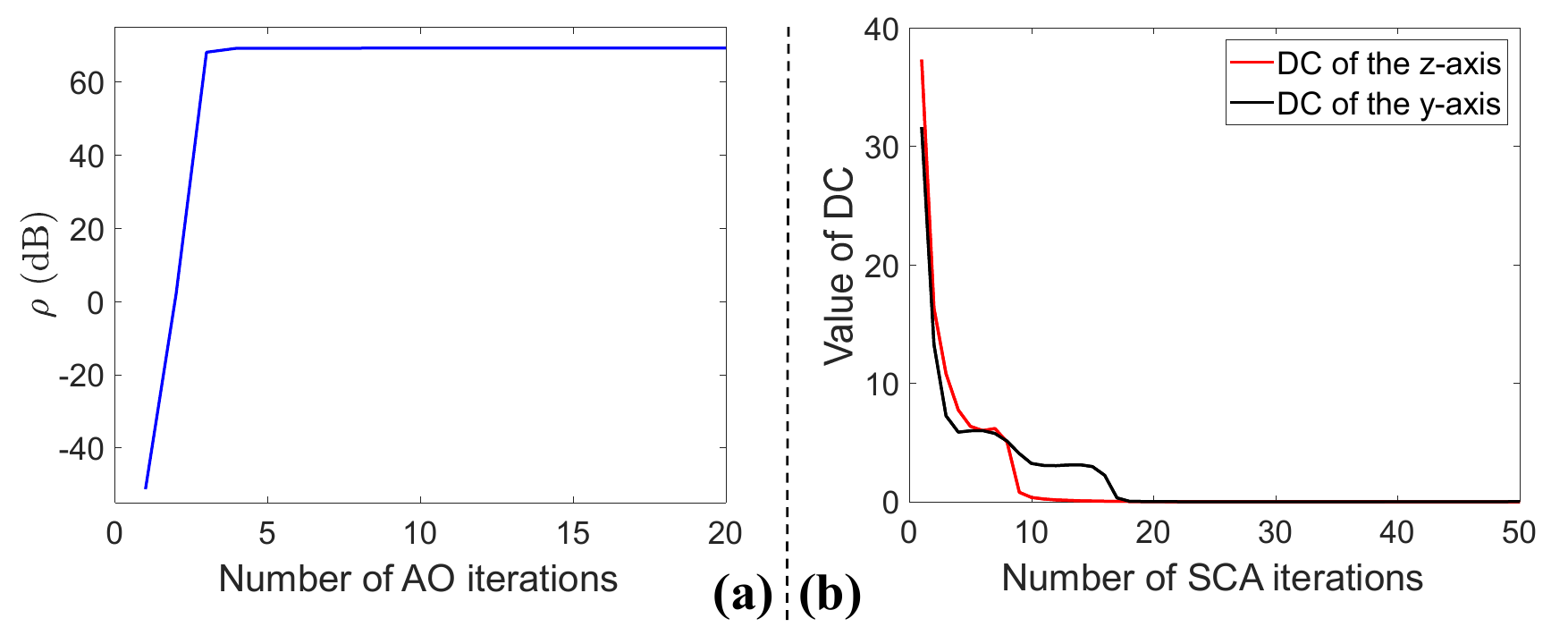}
	\caption{Convergence of (a) the AO procedure; and (b) SCA procedure.\vspace{-2ex}}
    \label{ConvergenceAnalysis}
\end{figure}

Finally, the convergence of the proposed algorithm is verified by simulation results demonstrated in Fig. \ref{ConvergenceAnalysis}.
One round of AO consists of two iterations each along the $y/z$-axis, respectively.
It can be seen that the DC component converges to zero within 20 iterations in the DC-SCA procedure, as shown in  Fig. \ref{ConvergenceAnalysis}(b). 
Furthermore, the overall power gain $\rho$ gets improved and converges in around 3-5 AO iterations as shown in Fig. \ref{ConvergenceAnalysis}(a).

\vspace{-2.5ex}



\subsection{Optimization of Large-Scale QS-IRS} 
\begin{figure} [htb]
	\centering
\includegraphics[width=0.65\linewidth,  trim=0 0 0 0,clip]{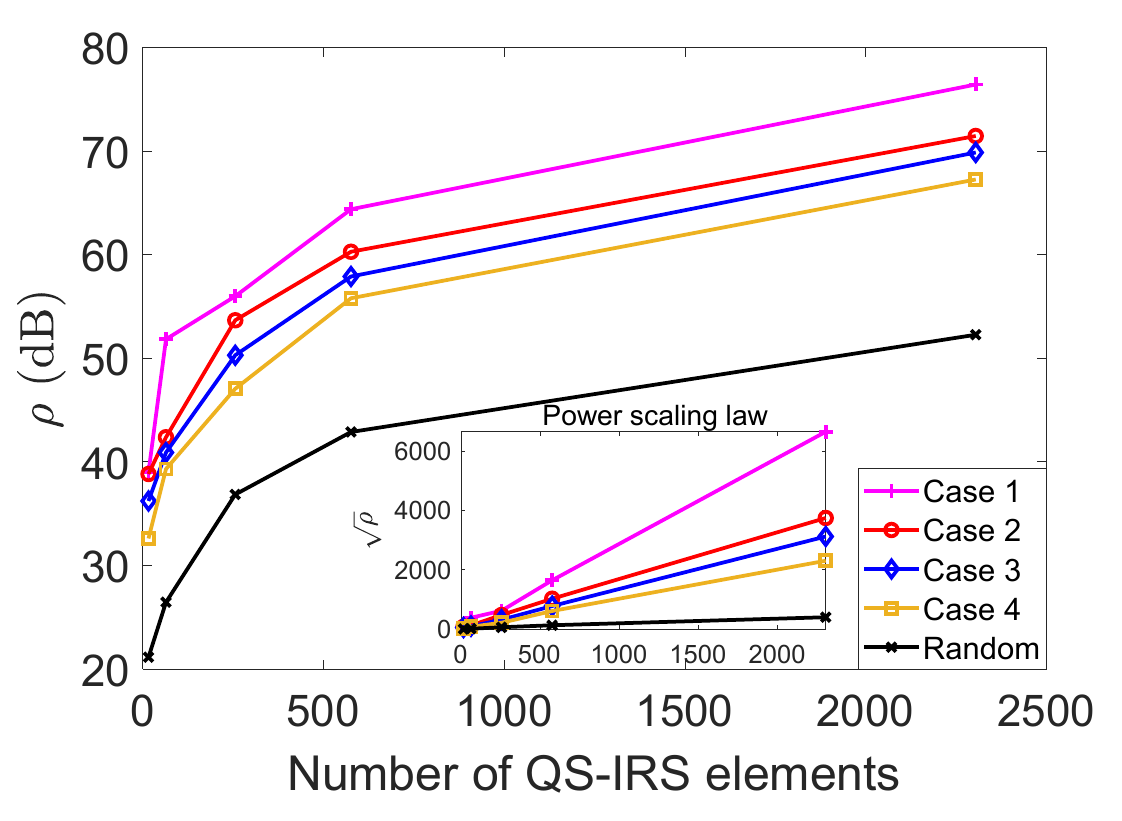}
	\caption{Power gain under different $M$ and mainlobe angular range: Case 1: $\varphi_p \in [15^{\circ},25^{\circ}]$, $\theta_p \in [125^{\circ},135^{\circ}]$; Case 2: $\varphi_p \in [10^{\circ},30^{\circ}]$, $\theta_p \in [120^{\circ},140^{\circ}]$; Case 3: $\varphi_p \in [0^{\circ},40^{\circ}]$, $\theta_p \in [120^{\circ},140^{\circ}]$; and Case 4: $\varphi_p \in [-10^{\circ},50^{\circ}]$, $\theta_p \in [120^{\circ},140^{\circ}]$.} \vspace{-2ex}\label{Number}
\end{figure}

We first compare AO with joint optimization under different number $M$ of elements, as in TABLE \ref{TableCompare}.
For small/moderate $M$, the power gain achieved by AO is close to that of the joint optimization, yet with significantly reduced running time.
As $M$ further increases, the joint optimization method runs prohibitively slow while AO is still applicable for large-scale QS-IRS (e.g., $M=48 \times 48$)\footnote{For $f_\textrm{c}=3.5$ GHz and half-wavelength element spacing, it may comprise $4\times 4$ panels each with $12\times 12$ elements and $0.5\times 0.5$ m$^2$ in size.} under reasonable running time, for applications like network planning and long-term coverage enhancement by QS-IRS.
Next, the power gains achieved by AO are plotted in Fig. \ref{Number}, where cases 1 to 4 represent co-centered mainlobe zones with increasing angular range.
Due to energy conservation, the power gain decreases as the area gradually expands, which is still significantly larger than that of the random-phase baseline.
Furthermore, it is observed from the subplot in Fig. \ref{Number} that the amplitude $\sqrt{\rho}$ grows approximately linearly with $M$, thus suggesting a power scaling law of $O(M^2)$ for our considered shaped beamforming design.

\begin{table}[htb]
\footnotesize
\centering
\caption{Comparisons between the joint optimization method and AO under $\varphi_p \in [-10^\circ, 10^\circ]$, $\theta_p \in [120^\circ, 140^\circ]$, and $\delta = 5$~dB.}
\addtolength{\tabcolsep}{-2pt}
\renewcommand{\arraystretch}{1.1}
\begin{tabular}{|c|c|c|c|c|c|c|}
\hline 
\multicolumn{2}{|c|}{Aperture size} & $4\times4$ & $8\times8$ & $16\times16$ & $24\times24$ & $48\times48$ \\
\hline
\multirow{2}{*}{\shortstack{Joint\\optimization}} & time(min) & 0.14 & 53.15 & - & - & - \\
\hhline{~------}
& $\rho$(dB) & 40.29 & 45.39 & - & - & - \\
\hline 
\multirow{2}{*}{AO} & time(min) & 0.22 & 0.39 & 1.44 & 5.19 & 133.95 \\
\hhline{~------}
& $\rho$(dB) & 40.29 & 45.01 & 52.82 & 60.01 & 69.73 \\
\hline 
\end{tabular}
\label{TableCompare}
\vspace{-2em}
\end{table}


\subsection{Power Gain Loss due to Phase Quantization} 
\begin{figure} [htb]
	\centering
	\includegraphics[width=0.6\linewidth,  trim=0 0 0 0,clip]{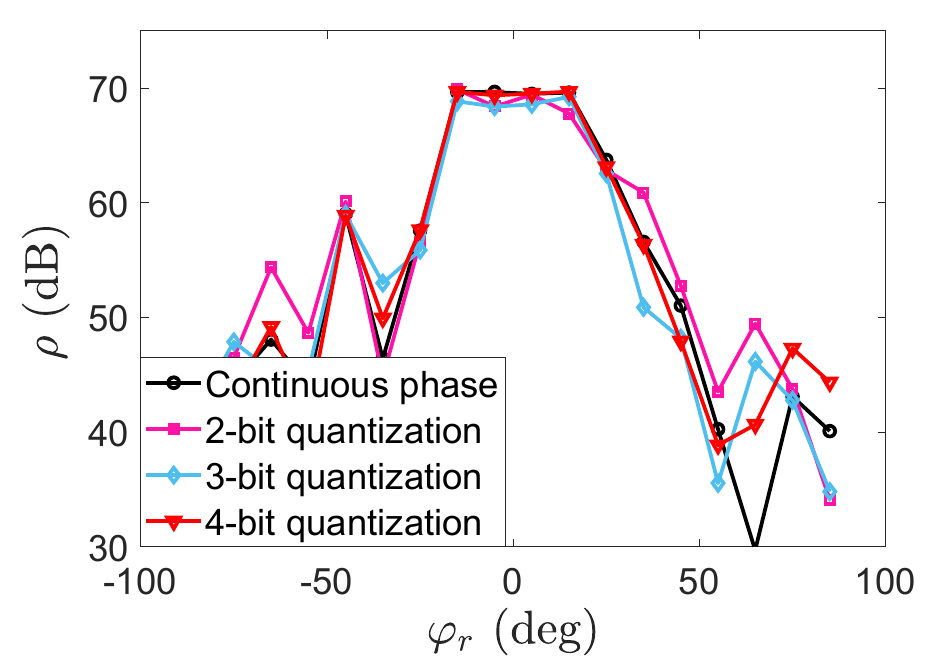}
	\caption{Power gain observed at $\theta_\text{r} = 140^{\circ}$.\vspace{-3ex}}\label{Quantization}
\end{figure}

Continuous phase is considered in the proposed algorithms. The power gain loss due to phase quantization is evaluated in Fig. \ref{Quantization}.
For the considered setup, the performance of 4-bit quantization is close to that of the continuous case, while 2-bit quantization brings about 2 dB loss. Practical applications need to trade-off between performance gain and complexity.

\vspace{-2ex}





\section{Conclusions and Future Work}
This paper introduces a new architecture of quasi-static IRS, which is suitable for low-cost and large-scale deployment to enhance long-term coverage.
In particular, the area coverage problem is formulated while considering IRS ERP, with the newly introduced shape masks for the mainlobe, and the sidelobe constraints to reduce energy leakage.
An AO algorithm based on the DC-SCA procedure is proposed, which achieves (non-rectangular/non-flat) shaped beamforming of the QS-IRS with power gain close to that of the joint optimization algorithm, and yet with significantly reduced complexity.
The impact of IRS deployment location and potential non-LoS channels will be considered in future work. Modular subarray pattern design is also interesting and worthy of further investigation.

\bibliography{IEEEabrv,bibliography}

\end{document}